\documentclass[
reprint,
superscriptaddress,
bibnotes,
 amsmath,amssymb, aps,
prl,
nolongbibliography
]{revtex4-2}
\usepackage{graphicx} 
\usepackage{dcolumn} 
\usepackage{bm} 
\usepackage[colorlinks,linkcolor=blue,citecolor=red,urlcolor=red]{hyperref}
\usepackage{nameref} 
\usepackage[utf8]{inputenc}
\usepackage{booktabs} 
\usepackage{float}
\usepackage{multirow}
\DeclareUnicodeCharacter{22C5}{\cdot}
\DeclareUnicodeCharacter{2212}{-}

\usepackage{color}
\usepackage{xcolor}

\begin{document}

\preprint{APS/123-QED}

\title{Hybrid and scalable photonic circuit cavity quantum electrodynamics}
\author{Xudong Wang}
\thanks{These authors contributed equally to this work}
\author{Yifan Zhu}
\thanks{These authors contributed equally to this work}
\author{Xiuqi Zhang}
\author{Yuanhao Qin}
\author{Bowen Chen}
\author{Yang Chen}
\affiliation{State Key Laboratory of Materials for Integrated Circuits, Shanghai Institute of Microsystem and Information Technology, Chinese Academy of Sciences, 865 Changning Road, Shanghai, 200050, China}
\affiliation{Center of Materials Science and Optoelectronics Engineering, University of Chinese Academy of Sciences, Beijing, 100049, China}
\author{Yongheng Huo}
\affiliation{Hefei National Research Center for Physical Sciences at the Microscale and School of physical Sciences, University of Science and Technology of China, Hefei, 230026, China}
\affiliation{Shanghai Research Center for Quantum Science and CAS Center for Excellence in Quantum Information and Quantum Physics, University of Science and Technology of China, Shanghai, 201315, China}
\author{Jiaxiang Zhang}
\email{jiaxiang.zhang@mail.sim.ac.cn}
\author{Xin Ou}
\email{ouxin@mail.sim.ac.cn}

\affiliation{State Key Laboratory of Materials for Integrated Circuits, Shanghai Institute of Microsystem and Information Technology, Chinese Academy of Sciences, 865 Changning Road, Shanghai, 200050, China}
\affiliation{Center of Materials Science and Optoelectronics Engineering, University of Chinese Academy of Sciences, Beijing, 100049, China}


\date{\today}

\begin{abstract}
Similar to superconducting circuit quantum electrodynamics (cQED), the development of a photonic analog—specifically, photonic circuit cQED—has become a major focus in integrated quantum photonics. Current solid-state cQED devices, however, face scalability challenges due to the difficulty in simultaneously spectral tuning of cavity modes and quantum emitters while ensuring in-plane optical modes confinement for efficient on-chip light routing. Here, we overcome these limitations by proposing and demonstrating a hybrid solid-state cQED platform integrated on a chip. Our device integrates semiconducting quantum dots (QDs) with a thin-film lithium niobate (TFLN) microring resonator. Leveraging TFLN’s ferroelectric and electro-optic (EO) properties, we implement local spectral tuning of both waveguide-coupled QDs and cavity modes. This approach achieves a broad spectral tuning range of up to 4.82 nm for individual QDs, enabling deterministic on-chip single-photon emission with a Purcell factor of 3.52. When combined with EO cavity tuning, we realize a spectrally tunable hybrid photonic circuit cQED device, sustaining near-constant Purcell factors of 1.89 over a 0.30 nm spectral range. This achievement enables scalable on-chip cavity-enhanced single-photon sources while preserving optical properties and maintaining compatibility with advanced photonic architectures, marking a significant step toward practical implementation of large-scale chip-based quantum networks.
\end{abstract}

\maketitle
\section*{Introduction}
Large-scale quantum simulators and networks require scalable platforms capable of integrating a large number of quantum bits (qubits)\cite{Beckman1996,Cirac2000}. Integrated quantum photonics (IQP) is one of the most promising modalities for this long-sought goal, offering unique ability to encode quantum information into photonic qubits with long coherence time and fast transmission speed\cite{Wang2019,Pelucchi2021}. While current demonstrations of chip-based photonic quantum applications rely mostly on parametric down-conversion sources, and they are probabilistic and thus problematic with scalability\cite{Politi2009}. Hybrid integration of solid-state quantum emitters (QEs) has emerged as a promising way to address the challenge, as highlighted by recent advancements in integrating various solid-state QEs into low-loss photonic circuits\cite{Elshaari2020,Kim2020,Wang2019,Li2024,chanana2022,Davanco2017}. Similar to the circuit cavity quantum electrodynamics on superconducting quantum platforms\cite{Blais2021,Blais2020}, developing a photonic analog is currently the focus of intensive research. The significance of such systems lies in their unique capability to enhance emission rates and thus protect qubits from decoherence\cite{Uppu2020,Katsumi2018,Zhu2025}, enabling deterministic single-photon generation with high brightness and indistinguishability. 

\begin{figure*}[t]
\includegraphics[width=0.8\textwidth]{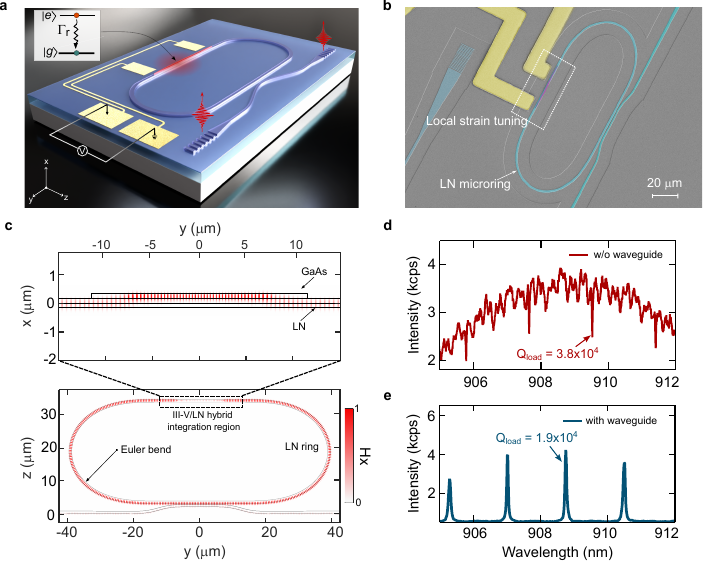}
\caption{\label{fig1:device sketch}\textbf{Hybrid photonic circuit cQED}.  (a) Schematic of the HPC-cQED. A tapered GaAs waveguide containing QDs is transfer-printed onto an $x$-cut TFLN racetrack resonator. The straight section of the resonator is oriented along the $z$-axis, with one side interfacing with a bus LN waveguide with grating couplers for optical input/output. A pair of metallic electrodes, spaced 5 $\mu m$ apart and situated 0.5 $\mu m$ from the racetrack section, generates an electric field along the opposing racetrack section, enabling localized strain fields to the hybrid GaAs/LN waveguide.  (\textbf{b}) Scanning electron microscope (SEM) image of the fabricated HPC-cQED device. Key components include the LN racetrack resonator, bus waveguide, grating couplers (blue), transferred GaAs taper (purple), and metallic electrodes (yellow). (\textbf{c}) Simulated WGMs in the hybrid GaAs/LN resonator. The taper-based mode converter efficiently confines the fundamental TE-like mode within the waveguide plane. Transmission spectrum of (\textbf{d}) the pure LN racetrack resonator and (\textbf{e}) the hybrid GaAs/LN resonator.}
\end{figure*}

Demonstrating photonic circuit cQED for scalable integrated quantum photonics is challenging. Current solid-state cQED devices are mostly implemented with nanophotonic cavities such as micropillar\cite{Wang2019dot,ding2016}, circular Bragg\cite{Sapienza2015,liu2019solid}, and open Fabry-Parot cavities\cite{ding2025,Najer2019,Kaupp2016,Riedel2017}. These cavities confine light in out-of-plane modes and thus are not suitable for on-chip integration. Moreover, the random formation of solid-state QEs and the microscopic imperfections in nanophotonic cavities create additional challenges for the scalable development of cQED on a chip\cite{Kim2016TPI}. The former arises from variations in the local electromagnetic environment of solid-state materials, which lead to inhomogeneous broadening of QEs photon emissions\cite{Grim2019,wang2022}. The latter stems from unavoidable fabrication variations in nanophotonic cavities, resulting in spectral mismatch between individual cQED devices. Demonstration of scalable photonic circuit cQED requires: (1) nanophotonic cavity with in-plane light confinement for efficient photonic state manipulation and light routing,  local and dynamic spectral tuning of (2) QEs and (3) cavity modes to enable quantum links in distributed nodes, and (4) large-scale, low-loss photonic circuits for integrating linear optical components on a single chip unit. 

The realization of scalable photonic circuit cQED devices has thus far prompted significant research efforts into planar nanophotonic cavities, spectral tuning methods for both QEs and cavity modes.  Systems that show high potential for IQP are two-dimensional planar photonic crystal (PhC) cavities with in-plane light confinement, especially those working in the slow-light regime\cite{Uppu2020,Uppu2021}. However, all PhC cavities reported have been implemented on monolithically integrated GaAs material platform, which suffers from high propagation losses at near-infrared wavelengths\cite{Papon2019}. Recently, there has been a thriving research momentum in developing one-dimensional nanobeam\cite{Katsumi2018,Davanco2017} and microring resonator cavities on photonic chips\cite{Dusanowski2020,Zhu2025}. However, scalable implementation of such cQED devices remains impractical due to the lack of simultaneous control over both the cavity and QEs.  Regarding the progress in cavity tuning, cryogenic gas condensation combined with microheating using coherent laser radiation remains the primary method\cite{Barclay2011,Li2024,Kim2016}. When extending to multiple cQED devices, however, each cavity must be controlled with an independent tunable laser, which turns out to be not scalable. As for spectral tuning for chip-integrated quantum emitters,  various \textit{post}-growth tuning techniques have been reported\cite{Wangx2022}, such as strain field based on bulk piezoelectric substrate\cite{jin2022generation,elshaari2018strain} and thermal-optical tuning\cite{Zhu2025,Osada2019}. However, these methods are either global or hampered by high power consumption. Although chip-based local spectral tuning to QEs has recently been demonstrated through either laser-induced phase transitions of a crystalline material\cite{Grim2019} or capacitive-induced strain fields\cite{Wang2019}, one caveat is that these methods are limited to compensating only the spectral difference of solid-state QEs, and there is no viable solution for manipulating the second degree of freedom. 

In this letter, we propose and demonstrate a hybrid photonic circuit cQED (HPC-cQED) device to address these challenges. Our approach leverages epitaxially grown III-V semiconductor QDs, which are well-established solid-state deterministic QEs, integrated onto a TFLN microring resonator with planar light confinement. Micro-photoluminescence ($\mu$-PL) measurements at cryogenic temperatures reveal that the HPC-cQED device supports whispering gallery modes (WGMs) with high quality factors up to 1.9$\times$10$^4$. By exploring the unique ferroelectric and EO properties of the thin-film LN, we introduce local and independent spectral tuning methods for the waveguide-coupled QDs and the ring cavity modes, respectively. This enables a broad and precise local spectral tuning range of up to 4.82 nm for individual QDs, achieving on-chip deterministic single-photon emission with a Purcell factor of 3.52. When further combined with EO spectral tuning for the cavity, we demonstrate a spectrally tunable HPC-cQED device with near-constant Purcell factors above 1.89 over a spectral range of 0.30 nm. This achievement allows demonstration of on-chip scalable cavity-enhanced single-photon sources while preserving optical properties and maintaining compatibility with advanced photonic architectures. Our approach represents a significant  milestone toward realizing large-scale chip-based quantum networks. 

\begin{figure*}[ht]
\includegraphics[width=1.0\textwidth]{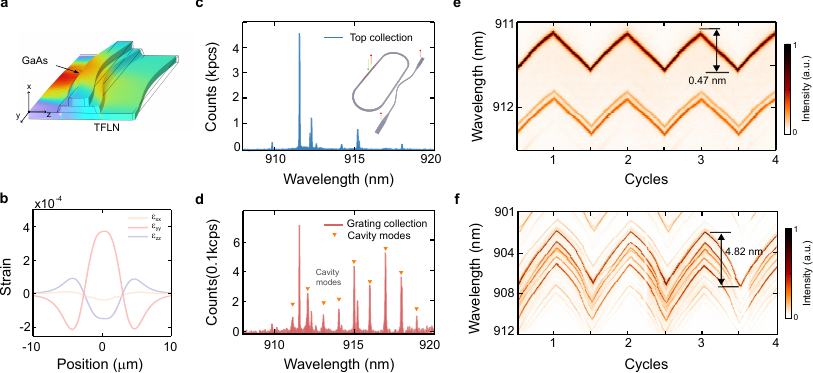}
\caption{\label{fig2:basic characterization}\textbf{On-chip local strain tuning of the waveguide-coupled QDs} (\textbf{a}) Finite element simulations of electric-field-induced displacements in the hybrid GaAs/LN waveguide. An voltage V$_{\mathrm{p}}$ = 500 V is applied along the $z$-axis of the $x$-cut TFLN. (\textbf{b}) Spatial profile of the principlal strain components ($\mathcal{E}_{\mathrm{xx}}$, $\mathcal{E}_{\mathrm{xx}}$ and $\mathcal{E}_{\mathrm{xx}}$)  of the cross-sectional hybrid GaAs/LN waveguide. $x$=0 represents the central position of the hybrid waveguide. Low temperature $\mu$-PL spectra of waveguide-coupled QDs collected from (\textbf{c}) the source position and (\textbf{d}) the grating coupler respectively. The inset shows the experimental arrangement of optical excitation (green) and PL light collection (red). The regular PL lines marked by the orange bars indicate the WGM modes of the hybrid GaAs/LN ring resonator. (\textbf{e}) Two-dimensional raster scan of photon emission from waveguide-coupled QDs under an repeated voltage from -500 to 500 \textit{V}. The PL is collected from the source position. (\textbf{f}) PL map of photon emission from the hyrbid GaAs/LN waveguide without the underlying SiO$_{\mathrm{2}}$ layer. The data was recorded by sweeping the voltage from -800 to 800 \textit{V}.}  
\end{figure*}

\section*{Results}
\subsection*{Design of the hybrid photonic circuit cQED}
The proposed HPC-cQED device consists of a GaAs waveguide integrated onto a high-quality factor racetrack ring resonator made of $x$-cut thin-film LN. A schematic diagram of the device is shown in Fig.\ref{fig1:device sketch}a. The QEs used here are InGaAs QDs embedded in the middle of a 180 nm thick GaAs layer grown by solid-source molecular beam epitaxy. An all-dry transfer method with high positioning accuracy was employed to fabricate the hybrid GaAs/LN with hundreds of nanometer precision. A scanning electron microscope image of the fabricated HPC-cQED device is shown in Fig. \ref{fig1:device sketch}b.  The geometric sizes of the superimposed GaAs waveguide and LN racetrack microring resonator are carefully designed so that the fundamental TE-like mode is preferentially supported at the central emission wavelength of QDs ($\sim 910$nm).  The GaAs waveguide is terminated with taper-based mode converters at both ends, enabling efficient coupling of  QDs photon emission from the overlaid GaAs layer to the underlying  LN waveguide \textit{via} evanescent wave coupling. Unlike uniform mode confinement in earlier hybrid ring resonators  \cite{Davanco2017, Dusanowski2020}, our hybrid ring cavity features a three-dimensional stack in which lossy GaAs waveguide covers a small portion of the LN ring resonator. This novel configuration reduces light propagation losses by minimizing GaAs-related scattering and absorption.  Furthermore, the LN racetrack microresonator incorporates Euler bends which feature a radius that adiabatically transitions from infinity in the straight sections to a finite value in the curved sections (see Supplementary Note 1 for more details). This design achieves low-loss planar resonators while maintaining a compact footprint, which is essential both for minimizing the mode volume of the entire QED system and for realizing WGMs in the hybrid ring structure (see Fig. \ref{fig1:device sketch}c) 

Although the intrinsic absorption loss from the material is suppressed, integration of the externally transferred GaAs waveguide introduces additional loss due to the mode transformer. The impact of this loss on the critical coupling quality factor (\textit{Q})—the key figure of merit for the hybrid cQED—can be modeled as:  Q = $\pi$\textit{n}$_{\mathrm{g}}$/$\lambda(\alpha_{\mathrm{total}})$. Where $\alpha_{\mathrm{total}} \approx (\alpha_{\mathrm{GaAs}}L_{\mathrm{GaAs}}+\alpha_{\mathrm{M}}L_{\mathrm{M}})/L_{\mathrm{total}}$ , \textit{n}$_{\mathrm{g}}$ represents  group index of the ring waveguide , and $\alpha_{\text{M}}$ quantifies the coupling loss of the mode transformer. Previous literature reports $\alpha_{\mathrm{GaAs}}$ as 75 dB$\cdot$cm$^{-1}$ at the near infrared wavelengths\cite{Papon2019} .  The coupling loss $\alpha_{\text{M}}$  is related to the mode transformer efficiency $\eta$ by:  $\alpha_{\text{M}}$  =-10$\times$log$_{\text{10}}\eta/L_{\text{M}}$, with $L_{\text{M}}$ being the length of the mode transformer. For a 10.5 $\mu$m-long taper,  $\eta$ =98.2$\%$ and $\alpha_{\text{M}}$ = 28.5 dB$\cdot$cm$^{-1}$ (see Supplementary note 1). This significant loss suggests that the hybrid cavity's Q factor will be substantially lower compared to the ring resonator without the GaAs waveguide.  Experimental characterization of the hybrid microring resonator was performed using a supercontinuum laser coupled via grating couplers. Transmission measurements (Fig.\ref{fig1:device sketch}d) reveal Q values of  4.2$\times$10$^4$ for the pure LN ring resonator. This lower Q value, limited by the system’s spectral resolution ($\sim$20 pm, see optical setup in Supplementary note 4), represents a lower bound for the  pure LN ring resonator with ultra-small radius. The suppressed light extinction ratio in the transmission curve implies that the pure LN resonator operates in overcoupling regime. Introducing the GaAs taper waveguide is expected to shift the system toward critical coupling by increasing propagation loss.  Fig.\ref{fig1:device sketch}e shows the transmission curve of the hybrid GaAs/LN ring resonator. The resulting resonant peaks indicate Q values of approximately 1.9$\times$10$^4$. Then we use \textit{V}$_{\mathrm{eff}}$ = $\int_{V}$$d^3\mathcal{r}$ $\epsilon$(\textbf{r}) $\mid$\textbf{E}(\textbf{r}) $\mid$$^2$/max$\lbrace$$\epsilon$(\textbf{r})  $\mid$\textbf{E}(\textbf{r}) $\mid$$^2$$\rbrace$  to determine the effective mode volume of the hybrid cavity. By employing a three-dimensional frequency-time domain time difference simulation method, the effective mode volume is found to be \textit{V}$_{\mathrm{eff}}$= 96.4 $(\lambda/n)^3$ (\textit{n} is the GaAs refractive index surrounding the QDs). Consequently, a maximum Purcell factor \textit{F}$_{\mathrm{p}}\sim$ 15 for our hybrid microcavity can be expected. 

\subsection*{Locally strain tuning to waveguide-coupled QDs}

We now introduce a local spectral tuning method to control the optical properties of waveguide-coupled QDs by exploiting the ferroelectric property of the TFLN. When a d.c. voltage (\textit{$V$}$_{\mathrm{p}}$) is applied along the $z$ axis of the $x$-cut TFLN, mesoscopic deformation can be generated, as shown in Fig. \ref{fig2:basic characterization}a. The pronounced structure displacement indicates the formation of local strain fields at the hybrid GaAs/LN waveguide. Given the piezoelectric coefficients of LN, the strain tensor ($\mathcal{E}$) in GaAs can be modeled as: 
\begin{equation}
		\label{eq1}
		\mathcal{E} = -\mathcal{S}_{\mathrm{E}}^{\mathrm{GaAs}}e^\mathcal{T}_{\mathrm{X-LN}}\mathcal{F}_{\mathrm{p}}
	\end{equation}
where $\mathcal{S}_{\mathrm{E}}^{\mathrm{GaAs}}$ is the compliance matrix of GaAs and $e_{\mathrm{X-LN}}$ denotes the piezoelectric constant of the $x$-cut LN. The piezoelectric constant $e^{\mathcal{T}}_{\mathrm{X-LN}}$ is related to the $\mathcal{Z}$-cut LN piezoelectric constant $e^{\mathcal{T}}_{\mathrm{\mathcal{Z}-LN}}$ ,through the rotation matrix $A$ and the stress tensor bond transformation matrix \textit{$\mathcal{M}$}, such that $e^{\mathcal{T}}_{\mathrm{X-LN}}$ = $A e^T_{\mathrm{\mathcal{Z}-LN}}\mathcal{M}^{\mathcal{T}}$. For a given voltage (\textit{V}$_{\mathrm{p}}$= 500 V), the dominant strain tensor components ($\mathcal{E}_{\mathrm{xx}}$, $\mathcal{E}_{\mathrm{yy}}$ and $\mathcal{E}_{\mathrm{zz}}$) are calculated and visualized in Fig. \ref{fig2:basic characterization}b.  Theoretically, we evaluated the impact of this strain tensor on waveguide-coupled QDs by calculating the GaAs band gap with the Pikus-Bir Hamiltonian\cite{Sun2010}, yielding $\Delta E_{\mathrm{GaAs}} = (a_{\mathrm{c}}+a_{\mathrm{v}})\mathcal{E}_{\mathrm{h}} - \sqrt{|\mathcal{Q}_{\mathcal{E}}|^2+|\mathcal{R}_{\mathcal{E}}|^2}$, with $\mathcal{E}_{\mathrm{h}} = \mathcal{E}_{\mathrm{xx}}+ \mathcal{E}_{\mathrm{yy}}+ \mathcal{E}_{\mathrm{zz}}$ being the hydrostatic strain. $\mathcal{Q}_{\mathcal{E}}=-b/2(\mathcal{E}_{\mathrm{xx}}+ \mathcal{E}_{\mathrm{yy}}-2\mathcal{E}_{\mathrm{zz}})$, $\mathcal{R}_{\mathcal{E}}=\sqrt{3}/2 b(\mathcal{E}_{\mathrm{xx}}-\mathcal{E}_{\mathrm{yy}})-id\mathcal{E}_{\mathrm{xy}}$, and $a_{\mathrm{c}}$, $a_{\mathrm{v}}$, $b$, $d$ are the deformation potentials of GaAs. This theory predicts a linear change in the GaAs band gap in response to $\mathcal{E}$ (see Supplementary Note 2 for more details). 

\begin{figure}[ht]
\includegraphics[width=0.48\textwidth]{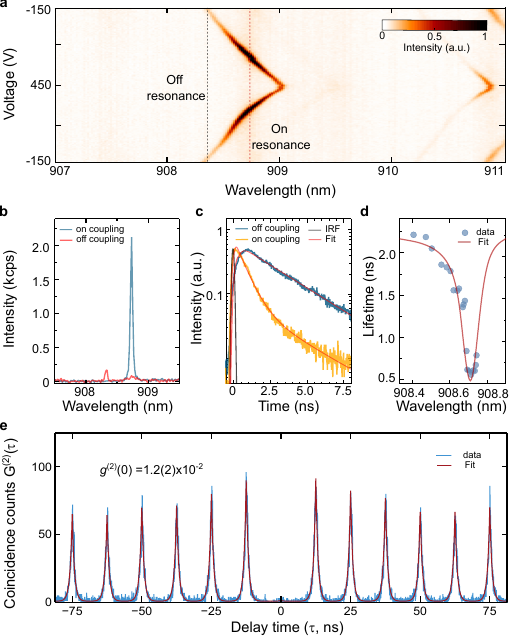}
\caption{\label{fig3:lifetimemeasurement}\textbf{Chip-integrated HPC-cQED for Purcell-enhanced single-photon emission} (\textbf{a}) On-chip local strain tuning of a single QD emission coupled to the hybrid GaAs/LN ring resonator. Dashed lines mark the QD emission at on-resonance (red) and off-resonance (black) with the cavity mode.  (b) Spectra of strain-tuned QD emission. A $\sim$10-fold intensity enhancement occurs when the QD is resonantly coupled to the cavity mode (red) compared to the off-resonant case (black).  (\textbf{c}) Time-resolved PL decay curves of QD emission at different detuning wavelengths. Solid lines: exponential decay fits accounting for the instrument response function (gray, $\tau_{\mathrm{IRF}}\sim$99.3 ps). (\textbf{d}) Lifetime as a function of detuning wavelengths and the solid line is a Lorentzian fit. (\textbf{e}) Second-order autocorrelation measurement of cavity-enhanced QD emission. }
\end{figure}

To experimentally validate the local strain fields in our HPC-cQED device, low-temperature $\mu$-PL measurements were carried out. QDs embedded in the GaAs tapered waveguide were optically excited using a 785 nm continuous-wave laser at the top (green arrow, inset), and photon emission was either collected from the top position or via the grating couplers (red arrows). The resulting spectra are shown in Fig. \ref{fig2:basic characterization}c and Fig. \ref{fig2:basic characterization}d, respectively. A dominant peak at 912.3 nm exhibits identical peak positions and spectral line widths across both collection configurations. This observation reveals efficient photon routing of a single QD in the hybrid GaAs/LN structure, mediated by the ring resonator and bus waveguide, and ultimately scattered by the grating couplers. In contrast to the spectrum obtained from the top position, the spectrum collected from the grating coupler exhibits discrete and regularly spaced spectral lines. These peaks arise from the fundamental WGMs of the hybrid cavity, with the free spectral range (FSR) being measured to be about 1.83 nm, which is consistent with the transmission curve shown in Fig. \ref{fig1:device sketch}e. Notably, these modes are exclusively detectable from the grating coupler, confirming strong in-plane light confinement in the hybrid cavity.  Next, a voltage V$_{\mathrm{p}}$ (with associated electric field $\mathcal{F}_{\mathrm{p}}$) was applied to the electrode in order to exert local strain fields to tune QD emission. Fig. \ref{fig2:basic characterization}e shows the spectral shift as \textit{V}$_{\mathrm{p}}$ is swept from -500 V to 500 V. A blue shift of $\Delta\lambda$ = 0.47 nm was observed with a tuning rate of 0.47 pm$\cdot$V$^{−1}$. The QD emission wavelength exhibits a linear dependence on \textit{V}$_{\mathrm{p}}$, unequivocally excluding the electric-field-induced lateral quantum Stark effect , which typically results in a quadratic wavelength tuning characteristic. To maximize the local strain tuning capability, we suspended the hybrid GaAs/LN waveguide through undercutting the bottom 4.7 $\mu$m thick SiO$_2$ layer. Fig. \ref{fig2:basic characterization}f demonstrates the improved performance of the device.  A maximum wavelength shift of 4.82 nm was achieved by sweeping \textit{V}$_{\mathrm{p}}$ from -800 V to 800 V, yielding a tuning rate (3.01 pm$\cdot$\textit{V}$^{−1}$),  sixfold that of the device with the SiO$_2$ layer. Multiple voltage cycling demonstrated consistent reproducibility of the maximum spectral tuning range (Fig. \ref{fig2:basic characterization}f), with no significant change in the QD emission linewidth. This highlights the excellent repeatability and stability of our tuning method.

\subsection*{On-chip cavity-enhanced single-photon emission and scalable cQED}
We now demonstrate the deterministic enhancement of single-photon emission from waveguide-coupled QDs in the hybrid cavity. Figure \ref{fig3:lifetimemeasurement}a shows PL spectra of a single QD collected via the grating coupler, with the emission wavelength being dynamically tuned into resonance with the cavity mode. A two-dimensional raster scan reveals a pronounced enhancement of photon emission when the QD is in resonance with the cavity mode at $\lambda_{\mathrm{c}}$ = 908.75 nm. The photon emission intensity increases tenfold compared to the off-resonant case ($\Delta\lambda$=0.41 nm), as shown in Fig. \ref{fig3:lifetimemeasurement}b.  To quantitatively evaluate the enhancement related to the Purcell effect, time-resolved PL measurements were performed. Fig. \ref{fig3:lifetimemeasurement}c displays the lifetime results for off- and on-resonance conditions. Deconvoluted lifetime fits show an increase of the spontaneous emission rate from $\Gamma_{\mathrm{off}}$= $\tau_{\mathrm{off}}$$^{-1}$ = 0.42 $\pm$ 0.02 ns$^{-1}$ ($\Delta\lambda$ = 0.41 nm) to $\Gamma_{\mathrm{on}}$= $\tau_{\mathrm{on}}$$^{-1}$ = 1.90 $\pm$ 0.03 ns$^{-1}$ ($\Delta\lambda$= 0).  Fig. \ref{fig3:lifetimemeasurement}c shows a detailed measurement of the lifetime by varying the wavelength detuning more precisely, with the data being well-fitted by a Lorentzian peak function. Considering the spontaneous rate for on- and off-resonant cases, the Purcell factor is calculated to be $F_{\mathrm{p}}$ = $\Gamma_{\mathrm{on}}$/$\Gamma_{\mathrm{off}}$ -1= 3.52 $\pm$ 0.2. With this value, we can estimate the QD dipole radiation efficiency into the fundamental mode of the waveguide, known as the $\beta$-factor: $\beta$ = $F_{\mathrm{p}}$/(1+$F_{\mathrm{p}}$) = 77.8\%. Thereafter, time-resolved second-order autocorrelation measurements ($\mathcal{G}$$^{(2)}$($\tau$))  under pulsed  p-shell excitation at 899.2 nm were performed to characterize this cavity-enhanced QD non-classical photon emission. As shown in Fig. \ref{fig3:lifetimemeasurement}e, the histogram exhibits pronounced  antibunching at zero time delay. Periodic peaks spaced at the laser repetition interval of 12.5 ns are observed. A fit to the data yields a central peak suppression of 1.2$\pm$0.2\%, where the uncertainty represents one standard deviation obtained from a double-Gaussian fit to the central peak area. This result demonstrates near-unity purity of deterministic single-photon emission from this chip-integrated hybrid cQED device.
 
\begin{figure*}[bt]
\includegraphics[width=1.0\textwidth]{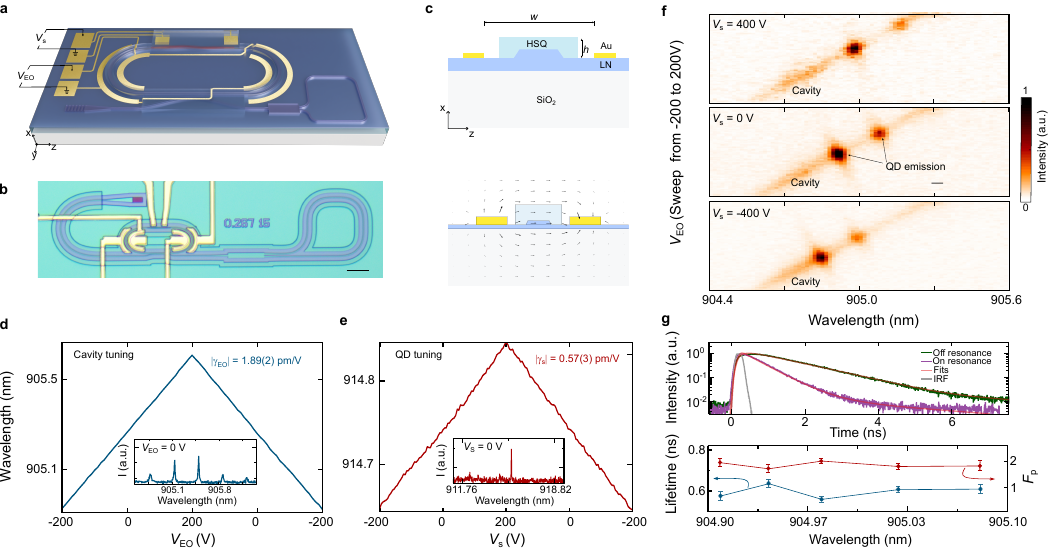}
\caption{\label{fig4:scalableoperation}\textbf{Chip-integrated scalable HPC-cQED} (\textbf{a}) Sketch of the modified HPC-cQED with additional electrode close to the Euler bends. (\textbf{b}) Microscopic image of the fabricated HPC-cQED which features two pairs of electrodes for EO-induced cavity tuning, one side pair of electrode for local strain tuning, a loop-mirror reflector and an output grating coupler. (\textbf{c}) Cross-sectional of the LN bending waveguide.  The geometric dimensions are: $w$=5 $\mu$m, $h$=800 nm. The bottom panel shows the simulated static electric field distribution (black arrows).  (\textbf{d}) Cavity resonance shift as a function of applied DC voltage (\textit{V}$_{\mathrm{EO}}$). The recorded resonance peak is shown in the inset. (\textbf{e}) QD emission shift as a function of applied DC voltage (\textit{V}$_{\mathrm{S}}$) and the inset shows the QD emission peak. (\textbf{f}) Wavelength-tunable QD emission coupled to the hybrid cavity mode engineered by the EO effect. (\textbf{g}) Time-resolved lifetime measurements at different QD-cavity resonant wavelengths. }
\end{figure*}

To enable scalable operation of our HPC-cQED device, a second tuning mechanism is needed to compensate spectral mismatches between different ring cavities caused by unavoidable fabrication imperfections. This is achieved in our studies by leveraging the large EO effect of LN to tune the hybrid ring resonator. We modified our design of the HPC-cQED device by placing two additional electrodes on both sides of the Euler bends. The sketch and fabricated device are shown in Figs. \ref{fig4:scalableoperation}a and b. To maximize the projection of the electric field parallel to the $z$-axis of LN (\textit{i.e.}, the \textit{c}-axis), metallic electrodes were directly deposited on both sides of the LN ridge waveguide with a gap of $w$=5 $\mu$m (see Fig. \ref{fig4:scalableoperation}c). A 800-nm-thick hydrogen silsesquioxane (HSQ) insulating layer was patterned and electrically cured on the waveguide to reduce optical losses from metal absorption\cite{lomonte2021}. Transmission measurements confirm that the hybrid microring resonator supports WGMs with Q-factors up to 1.2$\times$10$^4$. This reduced value, compared to the original design without electrodes and HSQ cover, is attributed to scattering by the HSQ layer and absorption due to additional complex electrodes. By applying independent voltages—V$_S$ (QD strain tuning) and V$_{EO}$ (cavity tuning)—we achieve linear spectral control on both the cavity mode and QD emission (Figs. \ref{fig4:scalableoperation}d and e). Over a voltage range from -200 to 200 V, the cavity mode shifts by 0.76 nm with a tuning rate of $\gamma_{\mathrm{EO}}$ = 1.89 $\pm$ 0.02 pm$\cdot$V$^{-1}$,  while the QD emission is shifted by 0.23 nm with a tuning rate of $\gamma_{\mathrm{S}}$ = 0.57 $\pm$ 0.03 pm$\cdot$V$^{-1}$. This dual-field tuning enables scalable operation of the hybrid cQED device: V$_S$ shifts the QD emission while V$_{EO}$ adjusts the cavity resonance, ensuring QD-cavity resonance over a spectral range of about 0.30 nm (see Fig. \ref{fig4:scalableoperation}f). To determine the Purcell enhancement at different tuning wavelengths, we performed time-resolved lifetime measurements on the cavity-coupled QD emission, as shown in Fig. \ref{fig4:scalableoperation}g.  Lifetimes for QD-cavity resonance are measured to be less than ($\tau_{\text{on}}\le$ 0.63 ns) at different wavelengths. When compared to the lifetime ($\tau_{\text{off}}$ = 1.82 ns) at large wavelength detuning,  Theoretical fits quantify that $F_{\mathrm{p}}$ = $\tau_{\text{off}}/\tau_{\text{on}}$-1 $>$ 1.89 over the entire tuning range of 0.30 nm, with V$_S$ varied from -200 to 200 V.  

\section*{Conclusion and outlook}
In conclusion, we have proposed and demonstrated a GaAs/LN hybrid cavity for chip-integrated cQED. A flexible transfer printing technique has been used to fabricate this HPC-cQED device through deterministically integrating a QD-containing GaAs waveguide onto a low-loss LN ring resonator. We explored the piezoelectric properties of TFLN and successfully introduced a circuit-compatible strain-tuning technique. This innovative method enables on-chip local, dynamic and reversible spectral tuning of individual QD emissions by up to 4.82 nm (7.30 meV at 910nm) — exceeding the transform-limited linewidth by three orders of magnitude\cite{Zhai2020}. This achievement facilitates the  demonstration of chip-integrated cavity-enhanced deterministic single-photon emission with a Purcell factor above 3.52. By further employing the EO spectral tuning for the cavity, we demonstrated a spectrally tunable HPC-cQED device with nearly constant Purcell factors above 1.89 over a spectral range of 0.30 nm. Such a scalable HPC-cQED device features strong in-plane light confinement, a dual-field spectral engineering method encompassing local strain and EO effect. Notably, the integrated and scalable HPC-QED systems, fabricated using a straightforward  transfer-printing technique, alleviate the complexity associated with photonic crystal-based integrated cavities. Critically, this method substantially mitigates reliance on bulky and intricate thermal-optic or gas condensation tuning methods for cavity control. This advancement marks a significant step toward realizing on-chip quantum interference between indistinguishable single photons from distant nodes in large-scale quantum networks.

The hybrid integrated quantum photonic chip is amenable to further optimizations for practical applications in large-scale chip-based quantum networks. For instance, the cavity performance can be further improved by optimizing the fabrication techniques so as to obtain higher Q factors. Possible directions toward this goal include improvement of the GaAs waveguide integration precision by using top-down wafer-bonding and nanofabrication techniques\cite{Davanco2017}, and suppression of absorption losses due to the complex metallic structures.  Additionally, active efforts are needed to pursue lifetime-limited single-photon emission from the device. A promising strategy at present is to apply resonant excitation\cite{Uppu2020}. All of these improvements presented here will open up the possibility to investigate rich physics in  chip-based all-solid-state QDs-coupled QED systems, paving the way for competitive integrated deterministic cavity-enhanced QEs for scalable quantum photonic applications.

\begin{acknowledgements}
The authors would like to acknowledge the financial support from the National Key R$\&$D Program of China (Grant No. 2022YFA1404604), the Chinese Academy of Sciences Project for Young Scientists in Basic Research (Grant No. YSBR-112), the National Natural Science Foundation of China (Grant Nos. 12074400, 62474168, 62293521, 62474168), the Strategic Priority Research Program of the Chinese Academy of Sciences (Grant No. XDB0670303), the Autonomous Deployment Project of the State Key Laboratory of Materials for Integrated Circuits (Grant No. SKLJC-Z2024-B03), and the State Key Laboratory of Advanced Optical Communication Systems and Networks (Grant No. 2024GZKF11). This work was carried out with the support of ShanghaiTech Material and Device Lab(SMDL20191219).
\end{acknowledgements}

\bibliography{reference}

\begin{thebibliography}{38}%
\makeatletter
\providecommand \@ifxundefined [1]{%
 \@ifx{#1\undefined}
}%
\providecommand \@ifnum [1]{%
 \ifnum #1\expandafter \@firstoftwo
 \else \expandafter \@secondoftwo
 \fi
}%
\providecommand \@ifx [1]{%
 \ifx #1\expandafter \@firstoftwo
 \else \expandafter \@secondoftwo
 \fi
}%
\providecommand \natexlab [1]{#1}%
\providecommand \enquote  [1]{``#1''}%
\providecommand \bibnamefont  [1]{#1}%
\providecommand \bibfnamefont [1]{#1}%
\providecommand \citenamefont [1]{#1}%
\providecommand \href@noop [0]{\@secondoftwo}%
\providecommand \href [0]{\begingroup \@sanitize@url \@href}%
\providecommand \@href[1]{\@@startlink{#1}\@@href}%
\providecommand \@@href[1]{\endgroup#1\@@endlink}%
\providecommand \@sanitize@url [0]{\catcode `\\12\catcode `\$12\catcode `\&12\catcode `\#12\catcode `\^12\catcode `\_12\catcode `\%12\relax}%
\providecommand \@@startlink[1]{}%
\providecommand \@@endlink[0]{}%
\providecommand \url  [0]{\begingroup\@sanitize@url \@url }%
\providecommand \@url [1]{\endgroup\@href {#1}{\urlprefix }}%
\providecommand \urlprefix  [0]{URL }%
\providecommand \Eprint [0]{\href }%
\providecommand \doibase [0]{https://doi.org/}%
\providecommand \selectlanguage [0]{\@gobble}%
\providecommand \bibinfo  [0]{\@secondoftwo}%
\providecommand \bibfield  [0]{\@secondoftwo}%
\providecommand \translation [1]{[#1]}%
\providecommand \BibitemOpen [0]{}%
\providecommand \bibitemStop [0]{}%
\providecommand \bibitemNoStop [0]{.\EOS\space}%
\providecommand \EOS [0]{\spacefactor3000\relax}%
\providecommand \BibitemShut  [1]{\csname bibitem#1\endcsname}%
\let\auto@bib@innerbib\@empty
\bibitem [{\citenamefont {Beckman}\ \emph {et~al.}(1996)\citenamefont {Beckman}, \citenamefont {Chari}, \citenamefont {Devabhaktuni},\ and\ \citenamefont {Preskill}}]{Beckman1996}%
  \BibitemOpen
  \bibfield  {author} {\bibinfo {author} {\bibfnamefont {D.}~\bibnamefont {Beckman}}, \bibinfo {author} {\bibfnamefont {A.~N.}\ \bibnamefont {Chari}}, \bibinfo {author} {\bibfnamefont {S.}~\bibnamefont {Devabhaktuni}},\ and\ \bibinfo {author} {\bibfnamefont {J.}~\bibnamefont {Preskill}},\ }\href {https://doi.org/10.1103/physreva.54.1034} {\bibfield  {journal} {\bibinfo  {journal} {Physical Review A}\ }\textbf {\bibinfo {volume} {54}},\ \bibinfo {pages} {1034–1063} (\bibinfo {year} {1996})}\BibitemShut {NoStop}%
\bibitem [{\citenamefont {Cirac}\ and\ \citenamefont {Zoller}(2000)}]{Cirac2000}%
  \BibitemOpen
  \bibfield  {author} {\bibinfo {author} {\bibfnamefont {J.~I.}\ \bibnamefont {Cirac}}\ and\ \bibinfo {author} {\bibfnamefont {P.}~\bibnamefont {Zoller}},\ }\href {https://doi.org/10.1038/35007021} {\bibfield  {journal} {\bibinfo  {journal} {Nature}\ }\textbf {\bibinfo {volume} {404}},\ \bibinfo {pages} {579–581} (\bibinfo {year} {2000})}\BibitemShut {NoStop}%
\bibitem [{\citenamefont {Wang}\ \emph {et~al.}(2019{\natexlab{a}})\citenamefont {Wang}, \citenamefont {Sciarrino}, \citenamefont {Laing},\ and\ \citenamefont {Thompson}}]{Wang2019}%
  \BibitemOpen
  \bibfield  {author} {\bibinfo {author} {\bibfnamefont {J.}~\bibnamefont {Wang}}, \bibinfo {author} {\bibfnamefont {F.}~\bibnamefont {Sciarrino}}, \bibinfo {author} {\bibfnamefont {A.}~\bibnamefont {Laing}},\ and\ \bibinfo {author} {\bibfnamefont {M.~G.}\ \bibnamefont {Thompson}},\ }\href {https://doi.org/10.1038/s41566-019-0532-1} {\bibfield  {journal} {\bibinfo  {journal} {Nature Photonics}\ }\textbf {\bibinfo {volume} {14}},\ \bibinfo {pages} {273–284} (\bibinfo {year} {2019}{\natexlab{a}})}\BibitemShut {NoStop}%
\bibitem [{\citenamefont {Pelucchi}\ \emph {et~al.}(2021)\citenamefont {Pelucchi}, \citenamefont {Fagas}, \citenamefont {Aharonovich}, \citenamefont {Englund}, \citenamefont {Figueroa}, \citenamefont {Gong}, \citenamefont {Hannes}, \citenamefont {Liu}, \citenamefont {Lu}, \citenamefont {Matsuda}, \citenamefont {Pan}, \citenamefont {Schreck}, \citenamefont {Sciarrino}, \citenamefont {Silberhorn}, \citenamefont {Wang},\ and\ \citenamefont {J\"{o}ns}}]{Pelucchi2021}%
  \BibitemOpen
  \bibfield  {author} {\bibinfo {author} {\bibfnamefont {E.}~\bibnamefont {Pelucchi}}, \bibinfo {author} {\bibfnamefont {G.}~\bibnamefont {Fagas}}, \bibinfo {author} {\bibfnamefont {I.}~\bibnamefont {Aharonovich}}, \bibinfo {author} {\bibfnamefont {D.}~\bibnamefont {Englund}}, \bibinfo {author} {\bibfnamefont {E.}~\bibnamefont {Figueroa}}, \bibinfo {author} {\bibfnamefont {Q.}~\bibnamefont {Gong}}, \bibinfo {author} {\bibfnamefont {H.}~\bibnamefont {Hannes}}, \bibinfo {author} {\bibfnamefont {J.}~\bibnamefont {Liu}}, \bibinfo {author} {\bibfnamefont {C.-Y.}\ \bibnamefont {Lu}}, \bibinfo {author} {\bibfnamefont {N.}~\bibnamefont {Matsuda}}, \bibinfo {author} {\bibfnamefont {J.-W.}\ \bibnamefont {Pan}}, \bibinfo {author} {\bibfnamefont {F.}~\bibnamefont {Schreck}}, \bibinfo {author} {\bibfnamefont {F.}~\bibnamefont {Sciarrino}}, \bibinfo {author} {\bibfnamefont {C.}~\bibnamefont {Silberhorn}}, \bibinfo {author} {\bibfnamefont {J.}~\bibnamefont {Wang}},\ and\ \bibinfo {author} {\bibfnamefont {K.~D.}\
  \bibnamefont {J\"{o}ns}},\ }\href {https://doi.org/10.1038/s42254-021-00398-z} {\bibfield  {journal} {\bibinfo  {journal} {Nature Reviews Physics}\ }\textbf {\bibinfo {volume} {4}},\ \bibinfo {pages} {194–208} (\bibinfo {year} {2021})}\BibitemShut {NoStop}%
\bibitem [{\citenamefont {Politi}\ \emph {et~al.}(2009)\citenamefont {Politi}, \citenamefont {Matthews}, \citenamefont {Thompson},\ and\ \citenamefont {O’Brien}}]{Politi2009}%
  \BibitemOpen
  \bibfield  {author} {\bibinfo {author} {\bibfnamefont {A.}~\bibnamefont {Politi}}, \bibinfo {author} {\bibfnamefont {J.}~\bibnamefont {Matthews}}, \bibinfo {author} {\bibfnamefont {M.}~\bibnamefont {Thompson}},\ and\ \bibinfo {author} {\bibfnamefont {J.}~\bibnamefont {O’Brien}},\ }\href {https://doi.org/10.1109/jstqe.2009.2026060} {\bibfield  {journal} {\bibinfo  {journal} {IEEE Journal of Selected Topics in Quantum Electronics}\ }\textbf {\bibinfo {volume} {15}},\ \bibinfo {pages} {1673–1684} (\bibinfo {year} {2009})}\BibitemShut {NoStop}%
\bibitem [{\citenamefont {Elshaari}\ \emph {et~al.}(2020)\citenamefont {Elshaari}, \citenamefont {Pernice}, \citenamefont {Srinivasan}, \citenamefont {Benson},\ and\ \citenamefont {Zwiller}}]{Elshaari2020}%
  \BibitemOpen
  \bibfield  {author} {\bibinfo {author} {\bibfnamefont {A.~W.}\ \bibnamefont {Elshaari}}, \bibinfo {author} {\bibfnamefont {W.}~\bibnamefont {Pernice}}, \bibinfo {author} {\bibfnamefont {K.}~\bibnamefont {Srinivasan}}, \bibinfo {author} {\bibfnamefont {O.}~\bibnamefont {Benson}},\ and\ \bibinfo {author} {\bibfnamefont {V.}~\bibnamefont {Zwiller}},\ }\href {https://doi.org/10.1038/s41566-020-0609-x} {\bibfield  {journal} {\bibinfo  {journal} {Nature Photonics}\ }\textbf {\bibinfo {volume} {14}},\ \bibinfo {pages} {285–298} (\bibinfo {year} {2020})}\BibitemShut {NoStop}%
\bibitem [{\citenamefont {Kim}\ \emph {et~al.}(2020)\citenamefont {Kim}, \citenamefont {Aghaeimeibodi}, \citenamefont {Carolan}, \citenamefont {Englund},\ and\ \citenamefont {Waks}}]{Kim2020}%
  \BibitemOpen
  \bibfield  {author} {\bibinfo {author} {\bibfnamefont {J.-H.}\ \bibnamefont {Kim}}, \bibinfo {author} {\bibfnamefont {S.}~\bibnamefont {Aghaeimeibodi}}, \bibinfo {author} {\bibfnamefont {J.}~\bibnamefont {Carolan}}, \bibinfo {author} {\bibfnamefont {D.}~\bibnamefont {Englund}},\ and\ \bibinfo {author} {\bibfnamefont {E.}~\bibnamefont {Waks}},\ }\href {https://doi.org/10.1364/optica.384118} {\bibfield  {journal} {\bibinfo  {journal} {Optica}\ }\textbf {\bibinfo {volume} {7}},\ \bibinfo {pages} {291} (\bibinfo {year} {2020})}\BibitemShut {NoStop}%
\bibitem [{\citenamefont {Li}\ \emph {et~al.}(2024)\citenamefont {Li}, \citenamefont {Santis}, \citenamefont {Harris}, \citenamefont {Chen}, \citenamefont {Gao}, \citenamefont {Christen}, \citenamefont {Choi}, \citenamefont {Trusheim}, \citenamefont {Song}, \citenamefont {Errando-Herranz}, \citenamefont {Du}, \citenamefont {Hu}, \citenamefont {Clark}, \citenamefont {Ibrahim}, \citenamefont {Gilbert}, \citenamefont {Han},\ and\ \citenamefont {Englund}}]{Li2024}%
  \BibitemOpen
  \bibfield  {author} {\bibinfo {author} {\bibfnamefont {L.}~\bibnamefont {Li}}, \bibinfo {author} {\bibfnamefont {L.~D.}\ \bibnamefont {Santis}}, \bibinfo {author} {\bibfnamefont {I.~B.~W.}\ \bibnamefont {Harris}}, \bibinfo {author} {\bibfnamefont {K.~C.}\ \bibnamefont {Chen}}, \bibinfo {author} {\bibfnamefont {Y.}~\bibnamefont {Gao}}, \bibinfo {author} {\bibfnamefont {I.}~\bibnamefont {Christen}}, \bibinfo {author} {\bibfnamefont {H.}~\bibnamefont {Choi}}, \bibinfo {author} {\bibfnamefont {M.}~\bibnamefont {Trusheim}}, \bibinfo {author} {\bibfnamefont {Y.}~\bibnamefont {Song}}, \bibinfo {author} {\bibfnamefont {C.}~\bibnamefont {Errando-Herranz}}, \bibinfo {author} {\bibfnamefont {J.}~\bibnamefont {Du}}, \bibinfo {author} {\bibfnamefont {Y.}~\bibnamefont {Hu}}, \bibinfo {author} {\bibfnamefont {G.}~\bibnamefont {Clark}}, \bibinfo {author} {\bibfnamefont {M.~I.}\ \bibnamefont {Ibrahim}}, \bibinfo {author} {\bibfnamefont {G.}~\bibnamefont {Gilbert}}, \bibinfo {author} {\bibfnamefont {R.}~\bibnamefont {Han}},\
  and\ \bibinfo {author} {\bibfnamefont {D.}~\bibnamefont {Englund}},\ }\href {https://doi.org/10.1038/s41586-024-07371-7} {\bibfield  {journal} {\bibinfo  {journal} {Nature}\ }\textbf {\bibinfo {volume} {630}},\ \bibinfo {pages} {70–76} (\bibinfo {year} {2024})}\BibitemShut {NoStop}%
\bibitem [{\citenamefont {Chanana}\ \emph {et~al.}(2022)\citenamefont {Chanana}, \citenamefont {Larocque}, \citenamefont {Moreira}, \citenamefont {Carolan}, \citenamefont {Guha}, \citenamefont {Melo}, \citenamefont {Anant}, \citenamefont {Song}, \citenamefont {Englund}, \citenamefont {Blumenthal}, \citenamefont {Srinivasan},\ and\ \citenamefont {Davanco}}]{chanana2022}%
  \BibitemOpen
  \bibfield  {author} {\bibinfo {author} {\bibfnamefont {A.}~\bibnamefont {Chanana}}, \bibinfo {author} {\bibfnamefont {H.}~\bibnamefont {Larocque}}, \bibinfo {author} {\bibfnamefont {R.}~\bibnamefont {Moreira}}, \bibinfo {author} {\bibfnamefont {J.}~\bibnamefont {Carolan}}, \bibinfo {author} {\bibfnamefont {B.}~\bibnamefont {Guha}}, \bibinfo {author} {\bibfnamefont {E.~G.}\ \bibnamefont {Melo}}, \bibinfo {author} {\bibfnamefont {V.}~\bibnamefont {Anant}}, \bibinfo {author} {\bibfnamefont {J.}~\bibnamefont {Song}}, \bibinfo {author} {\bibfnamefont {D.}~\bibnamefont {Englund}}, \bibinfo {author} {\bibfnamefont {D.~J.}\ \bibnamefont {Blumenthal}}, \bibinfo {author} {\bibfnamefont {K.}~\bibnamefont {Srinivasan}},\ and\ \bibinfo {author} {\bibfnamefont {M.}~\bibnamefont {Davanco}},\ }\href {https://doi.org/10.1038/s41467-022-35332-z} {\bibfield  {journal} {\bibinfo  {journal} {Nature Communications}\ }\textbf {\bibinfo {volume} {13}},\ \bibinfo {pages} {7693} (\bibinfo {year} {2022})}\BibitemShut {NoStop}%
\bibitem [{\citenamefont {Davanco}\ \emph {et~al.}(2017)\citenamefont {Davanco}, \citenamefont {Liu}, \citenamefont {Sapienza}, \citenamefont {Zhang}, \citenamefont {De~Miranda~Cardoso}, \citenamefont {Verma}, \citenamefont {Mirin}, \citenamefont {Nam}, \citenamefont {Liu},\ and\ \citenamefont {Srinivasan}}]{Davanco2017}%
  \BibitemOpen
  \bibfield  {author} {\bibinfo {author} {\bibfnamefont {M.}~\bibnamefont {Davanco}}, \bibinfo {author} {\bibfnamefont {J.}~\bibnamefont {Liu}}, \bibinfo {author} {\bibfnamefont {L.}~\bibnamefont {Sapienza}}, \bibinfo {author} {\bibfnamefont {C.-Z.}\ \bibnamefont {Zhang}}, \bibinfo {author} {\bibfnamefont {J.~V.}\ \bibnamefont {De~Miranda~Cardoso}}, \bibinfo {author} {\bibfnamefont {V.}~\bibnamefont {Verma}}, \bibinfo {author} {\bibfnamefont {R.}~\bibnamefont {Mirin}}, \bibinfo {author} {\bibfnamefont {S.~W.}\ \bibnamefont {Nam}}, \bibinfo {author} {\bibfnamefont {L.}~\bibnamefont {Liu}},\ and\ \bibinfo {author} {\bibfnamefont {K.}~\bibnamefont {Srinivasan}},\ }\href {https://doi.org/10.1038/s41467-017-00987-6} {\bibfield  {journal} {\bibinfo  {journal} {Nature Communications}\ }\textbf {\bibinfo {volume} {8}},\ \bibinfo {pages} {889} (\bibinfo {year} {2017})}\BibitemShut {NoStop}%
\bibitem [{\citenamefont {Blais}\ \emph {et~al.}(2021)\citenamefont {Blais}, \citenamefont {Grimsmo}, \citenamefont {Girvin},\ and\ \citenamefont {Wallraff}}]{Blais2021}%
  \BibitemOpen
  \bibfield  {author} {\bibinfo {author} {\bibfnamefont {A.}~\bibnamefont {Blais}}, \bibinfo {author} {\bibfnamefont {A.~L.}\ \bibnamefont {Grimsmo}}, \bibinfo {author} {\bibfnamefont {S.~M.}\ \bibnamefont {Girvin}},\ and\ \bibinfo {author} {\bibfnamefont {A.}~\bibnamefont {Wallraff}},\ }\href {https://doi.org/10.1103/RevModPhys.93.025005} {\bibfield  {journal} {\bibinfo  {journal} {Reviews of Modern Physics}\ }\textbf {\bibinfo {volume} {93}},\ \bibinfo {pages} {025005} (\bibinfo {year} {2021})}\BibitemShut {NoStop}%
\bibitem [{\citenamefont {Blais}\ \emph {et~al.}(2020)\citenamefont {Blais}, \citenamefont {Girvin},\ and\ \citenamefont {Oliver}}]{Blais2020}%
  \BibitemOpen
  \bibfield  {author} {\bibinfo {author} {\bibfnamefont {A.}~\bibnamefont {Blais}}, \bibinfo {author} {\bibfnamefont {S.~M.}\ \bibnamefont {Girvin}},\ and\ \bibinfo {author} {\bibfnamefont {W.~D.}\ \bibnamefont {Oliver}},\ }\href {https://doi.org/10.1038/s41567-020-0806-z} {\bibfield  {journal} {\bibinfo  {journal} {Nature Physics}\ }\textbf {\bibinfo {volume} {16}},\ \bibinfo {pages} {247–256} (\bibinfo {year} {2020})}\BibitemShut {NoStop}%
\bibitem [{\citenamefont {Uppu}\ \emph {et~al.}(2020)\citenamefont {Uppu}, \citenamefont {Pedersen}, \citenamefont {Wang}, \citenamefont {Olesen}, \citenamefont {Papon}, \citenamefont {Zhou}, \citenamefont {Midolo}, \citenamefont {Scholz}, \citenamefont {Wieck}, \citenamefont {Ludwig},\ and\ \citenamefont {Lodahl}}]{Uppu2020}%
  \BibitemOpen
  \bibfield  {author} {\bibinfo {author} {\bibfnamefont {R.}~\bibnamefont {Uppu}}, \bibinfo {author} {\bibfnamefont {F.~T.}\ \bibnamefont {Pedersen}}, \bibinfo {author} {\bibfnamefont {Y.}~\bibnamefont {Wang}}, \bibinfo {author} {\bibfnamefont {C.~T.}\ \bibnamefont {Olesen}}, \bibinfo {author} {\bibfnamefont {C.}~\bibnamefont {Papon}}, \bibinfo {author} {\bibfnamefont {X.}~\bibnamefont {Zhou}}, \bibinfo {author} {\bibfnamefont {L.}~\bibnamefont {Midolo}}, \bibinfo {author} {\bibfnamefont {S.}~\bibnamefont {Scholz}}, \bibinfo {author} {\bibfnamefont {A.~D.}\ \bibnamefont {Wieck}}, \bibinfo {author} {\bibfnamefont {A.}~\bibnamefont {Ludwig}},\ and\ \bibinfo {author} {\bibfnamefont {P.}~\bibnamefont {Lodahl}},\ }\href {https://doi.org/10.1126/sciadv.abc8268} {\bibfield  {journal} {\bibinfo  {journal} {Science Advances}\ }\textbf {\bibinfo {volume} {6}},\ \bibinfo {pages} {eabc8268} (\bibinfo {year} {2020})}\BibitemShut {NoStop}%
\bibitem [{\citenamefont {Katsumi}\ \emph {et~al.}(2018)\citenamefont {Katsumi}, \citenamefont {Ota}, \citenamefont {Kakuda}, \citenamefont {Iwamoto},\ and\ \citenamefont {Arakawa}}]{Katsumi2018}%
  \BibitemOpen
  \bibfield  {author} {\bibinfo {author} {\bibfnamefont {R.}~\bibnamefont {Katsumi}}, \bibinfo {author} {\bibfnamefont {Y.}~\bibnamefont {Ota}}, \bibinfo {author} {\bibfnamefont {M.}~\bibnamefont {Kakuda}}, \bibinfo {author} {\bibfnamefont {S.}~\bibnamefont {Iwamoto}},\ and\ \bibinfo {author} {\bibfnamefont {Y.}~\bibnamefont {Arakawa}},\ }\href {https://doi.org/10.1364/optica.5.000691} {\bibfield  {journal} {\bibinfo  {journal} {Optica}\ }\textbf {\bibinfo {volume} {5}},\ \bibinfo {pages} {691} (\bibinfo {year} {2018})}\BibitemShut {NoStop}%
\bibitem [{\citenamefont {Zhu}\ \emph {et~al.}(2025)\citenamefont {Zhu}, \citenamefont {Liu}, \citenamefont {Yi}, \citenamefont {Wang}, \citenamefont {Qin}, \citenamefont {Zhao}, \citenamefont {Zhao}, \citenamefont {Chen}, \citenamefont {Zhang}, \citenamefont {Song}, \citenamefont {Huo}, \citenamefont {Ou},\ and\ \citenamefont {Zhang}}]{Zhu2025}%
  \BibitemOpen
  \bibfield  {author} {\bibinfo {author} {\bibfnamefont {Y.}~\bibnamefont {Zhu}}, \bibinfo {author} {\bibfnamefont {R.}~\bibnamefont {Liu}}, \bibinfo {author} {\bibfnamefont {A.}~\bibnamefont {Yi}}, \bibinfo {author} {\bibfnamefont {X.}~\bibnamefont {Wang}}, \bibinfo {author} {\bibfnamefont {Y.}~\bibnamefont {Qin}}, \bibinfo {author} {\bibfnamefont {Z.}~\bibnamefont {Zhao}}, \bibinfo {author} {\bibfnamefont {J.}~\bibnamefont {Zhao}}, \bibinfo {author} {\bibfnamefont {B.}~\bibnamefont {Chen}}, \bibinfo {author} {\bibfnamefont {X.}~\bibnamefont {Zhang}}, \bibinfo {author} {\bibfnamefont {S.}~\bibnamefont {Song}}, \bibinfo {author} {\bibfnamefont {Y.}~\bibnamefont {Huo}}, \bibinfo {author} {\bibfnamefont {X.}~\bibnamefont {Ou}},\ and\ \bibinfo {author} {\bibfnamefont {J.}~\bibnamefont {Zhang}},\ }\href {https://doi.org/10.1038/s41377-024-01676-y} {\bibfield  {journal} {\bibinfo  {journal} {Light: Science \& Applications}\ }\textbf {\bibinfo {volume} {14}},\ \bibinfo {pages} {86} (\bibinfo {year}
  {2025})}\BibitemShut {NoStop}%
\bibitem [{\citenamefont {Wang}\ \emph {et~al.}(2019{\natexlab{b}})\citenamefont {Wang}, \citenamefont {He}, \citenamefont {Chung}, \citenamefont {Hu}, \citenamefont {Yu}, \citenamefont {Chen}, \citenamefont {Ding}, \citenamefont {Chen}, \citenamefont {Qin}, \citenamefont {Yang}, \citenamefont {Liu}, \citenamefont {Duan}, \citenamefont {Li}, \citenamefont {Gerhardt}, \citenamefont {Winkler}, \citenamefont {Jurkat}, \citenamefont {Wang}, \citenamefont {Gregersen}, \citenamefont {Huo}, \citenamefont {Dai}, \citenamefont {Yu}, \citenamefont {H{\"o}fling}, \citenamefont {Lu},\ and\ \citenamefont {Pan}}]{Wang2019dot}%
  \BibitemOpen
  \bibfield  {author} {\bibinfo {author} {\bibfnamefont {H.}~\bibnamefont {Wang}}, \bibinfo {author} {\bibfnamefont {Y.-M.}\ \bibnamefont {He}}, \bibinfo {author} {\bibfnamefont {T.-H.}\ \bibnamefont {Chung}}, \bibinfo {author} {\bibfnamefont {H.}~\bibnamefont {Hu}}, \bibinfo {author} {\bibfnamefont {Y.}~\bibnamefont {Yu}}, \bibinfo {author} {\bibfnamefont {S.}~\bibnamefont {Chen}}, \bibinfo {author} {\bibfnamefont {X.}~\bibnamefont {Ding}}, \bibinfo {author} {\bibfnamefont {M.-C.}\ \bibnamefont {Chen}}, \bibinfo {author} {\bibfnamefont {J.}~\bibnamefont {Qin}}, \bibinfo {author} {\bibfnamefont {X.}~\bibnamefont {Yang}}, \bibinfo {author} {\bibfnamefont {R.-Z.}\ \bibnamefont {Liu}}, \bibinfo {author} {\bibfnamefont {Z.-C.}\ \bibnamefont {Duan}}, \bibinfo {author} {\bibfnamefont {J.-P.}\ \bibnamefont {Li}}, \bibinfo {author} {\bibfnamefont {S.}~\bibnamefont {Gerhardt}}, \bibinfo {author} {\bibfnamefont {K.}~\bibnamefont {Winkler}}, \bibinfo {author} {\bibfnamefont {J.}~\bibnamefont {Jurkat}}, \bibinfo {author}
  {\bibfnamefont {L.-J.}\ \bibnamefont {Wang}}, \bibinfo {author} {\bibfnamefont {N.}~\bibnamefont {Gregersen}}, \bibinfo {author} {\bibfnamefont {Y.-H.}\ \bibnamefont {Huo}}, \bibinfo {author} {\bibfnamefont {Q.}~\bibnamefont {Dai}}, \bibinfo {author} {\bibfnamefont {S.}~\bibnamefont {Yu}}, \bibinfo {author} {\bibfnamefont {S.}~\bibnamefont {H{\"o}fling}}, \bibinfo {author} {\bibfnamefont {C.-Y.}\ \bibnamefont {Lu}},\ and\ \bibinfo {author} {\bibfnamefont {J.-W.}\ \bibnamefont {Pan}},\ }\href {https://doi.org/10.1038/s41566-019-0494-3} {\bibfield  {journal} {\bibinfo  {journal} {Nature Photonics}\ }\textbf {\bibinfo {volume} {13}},\ \bibinfo {pages} {770} (\bibinfo {year} {2019}{\natexlab{b}})}\BibitemShut {NoStop}%
\bibitem [{\citenamefont {Ding}\ \emph {et~al.}(2016)\citenamefont {Ding}, \citenamefont {He}, \citenamefont {Duan}, \citenamefont {Gregersen}, \citenamefont {Chen}, \citenamefont {Unsleber}, \citenamefont {Maier}, \citenamefont {Schneider}, \citenamefont {Kamp}, \citenamefont {H{\"o}fling}, \citenamefont {Lu},\ and\ \citenamefont {Pan}}]{ding2016}%
  \BibitemOpen
  \bibfield  {author} {\bibinfo {author} {\bibfnamefont {X.}~\bibnamefont {Ding}}, \bibinfo {author} {\bibfnamefont {Y.}~\bibnamefont {He}}, \bibinfo {author} {\bibfnamefont {Z.-C.}\ \bibnamefont {Duan}}, \bibinfo {author} {\bibfnamefont {N.}~\bibnamefont {Gregersen}}, \bibinfo {author} {\bibfnamefont {M.-C.}\ \bibnamefont {Chen}}, \bibinfo {author} {\bibfnamefont {S.}~\bibnamefont {Unsleber}}, \bibinfo {author} {\bibfnamefont {S.}~\bibnamefont {Maier}}, \bibinfo {author} {\bibfnamefont {C.}~\bibnamefont {Schneider}}, \bibinfo {author} {\bibfnamefont {M.}~\bibnamefont {Kamp}}, \bibinfo {author} {\bibfnamefont {S.}~\bibnamefont {H{\"o}fling}}, \bibinfo {author} {\bibfnamefont {C.-Y.}\ \bibnamefont {Lu}},\ and\ \bibinfo {author} {\bibfnamefont {J.-W.}\ \bibnamefont {Pan}},\ }\href {https://doi.org/10.1103/PhysRevLett.116.020401} {\bibfield  {journal} {\bibinfo  {journal} {Physical Review Letters}\ }\textbf {\bibinfo {volume} {116}},\ \bibinfo {pages} {020401} (\bibinfo {year} {2016})}\BibitemShut {NoStop}%
\bibitem [{\citenamefont {Sapienza}\ \emph {et~al.}(2015)\citenamefont {Sapienza}, \citenamefont {Davan{\c c}o}, \citenamefont {Badolato},\ and\ \citenamefont {Srinivasan}}]{Sapienza2015}%
  \BibitemOpen
  \bibfield  {author} {\bibinfo {author} {\bibfnamefont {L.}~\bibnamefont {Sapienza}}, \bibinfo {author} {\bibfnamefont {M.}~\bibnamefont {Davan{\c c}o}}, \bibinfo {author} {\bibfnamefont {A.}~\bibnamefont {Badolato}},\ and\ \bibinfo {author} {\bibfnamefont {K.}~\bibnamefont {Srinivasan}},\ }\href {https://doi.org/10.1038/ncomms8833} {\bibfield  {journal} {\bibinfo  {journal} {Nature Communications}\ }\textbf {\bibinfo {volume} {6}},\ \bibinfo {pages} {7833} (\bibinfo {year} {2015})}\BibitemShut {NoStop}%
\bibitem [{\citenamefont {Liu}\ \emph {et~al.}(2019)\citenamefont {Liu}, \citenamefont {Su}, \citenamefont {Wei}, \citenamefont {Yao}, \citenamefont {da~Silva}, \citenamefont {Yu}, \citenamefont {{Iles-Smith}}, \citenamefont {Srinivasan}, \citenamefont {Rastelli}, \citenamefont {Li},\ and\ \citenamefont {Wang}}]{liu2019solid}%
  \BibitemOpen
  \bibfield  {author} {\bibinfo {author} {\bibfnamefont {J.}~\bibnamefont {Liu}}, \bibinfo {author} {\bibfnamefont {R.}~\bibnamefont {Su}}, \bibinfo {author} {\bibfnamefont {Y.}~\bibnamefont {Wei}}, \bibinfo {author} {\bibfnamefont {B.}~\bibnamefont {Yao}}, \bibinfo {author} {\bibfnamefont {S.~F.~C.}\ \bibnamefont {da~Silva}}, \bibinfo {author} {\bibfnamefont {Y.}~\bibnamefont {Yu}}, \bibinfo {author} {\bibfnamefont {J.}~\bibnamefont {{Iles-Smith}}}, \bibinfo {author} {\bibfnamefont {K.}~\bibnamefont {Srinivasan}}, \bibinfo {author} {\bibfnamefont {A.}~\bibnamefont {Rastelli}}, \bibinfo {author} {\bibfnamefont {J.}~\bibnamefont {Li}},\ and\ \bibinfo {author} {\bibfnamefont {X.}~\bibnamefont {Wang}},\ }\href {https://doi.org/10.1038/s41565-019-0435-9} {\bibfield  {journal} {\bibinfo  {journal} {Nature Nanotechnology}\ }\textbf {\bibinfo {volume} {14}},\ \bibinfo {pages} {586} (\bibinfo {year} {2019})}\BibitemShut {NoStop}%
\bibitem [{\citenamefont {Ding}\ \emph {et~al.}(2025)\citenamefont {Ding}, \citenamefont {Guo}, \citenamefont {Xu}, \citenamefont {Liu}, \citenamefont {Zou}, \citenamefont {Zhao}, \citenamefont {Ge}, \citenamefont {Zhang}, \citenamefont {Liu}, \citenamefont {Wang}, \citenamefont {Chen}, \citenamefont {Wang}, \citenamefont {He}, \citenamefont {Huo}, \citenamefont {Lu},\ and\ \citenamefont {Pan}}]{ding2025}%
  \BibitemOpen
  \bibfield  {author} {\bibinfo {author} {\bibfnamefont {X.}~\bibnamefont {Ding}}, \bibinfo {author} {\bibfnamefont {Y.-P.}\ \bibnamefont {Guo}}, \bibinfo {author} {\bibfnamefont {M.-C.}\ \bibnamefont {Xu}}, \bibinfo {author} {\bibfnamefont {R.-Z.}\ \bibnamefont {Liu}}, \bibinfo {author} {\bibfnamefont {G.-Y.}\ \bibnamefont {Zou}}, \bibinfo {author} {\bibfnamefont {J.-Y.}\ \bibnamefont {Zhao}}, \bibinfo {author} {\bibfnamefont {Z.-X.}\ \bibnamefont {Ge}}, \bibinfo {author} {\bibfnamefont {Q.-H.}\ \bibnamefont {Zhang}}, \bibinfo {author} {\bibfnamefont {H.-L.}\ \bibnamefont {Liu}}, \bibinfo {author} {\bibfnamefont {L.-J.}\ \bibnamefont {Wang}}, \bibinfo {author} {\bibfnamefont {M.-C.}\ \bibnamefont {Chen}}, \bibinfo {author} {\bibfnamefont {H.}~\bibnamefont {Wang}}, \bibinfo {author} {\bibfnamefont {Y.-M.}\ \bibnamefont {He}}, \bibinfo {author} {\bibfnamefont {Y.-H.}\ \bibnamefont {Huo}}, \bibinfo {author} {\bibfnamefont {C.-Y.}\ \bibnamefont {Lu}},\ and\ \bibinfo {author} {\bibfnamefont {J.-W.}\ \bibnamefont
  {Pan}},\ }\href {https://doi.org/10.1038/s41566-025-01639-8} {\bibfield  {journal} {\bibinfo  {journal} {Nature Photonics}\ ,\ \bibinfo {pages} {1}} (\bibinfo {year} {2025})}\BibitemShut {NoStop}%
\bibitem [{\citenamefont {Najer}\ \emph {et~al.}(2019)\citenamefont {Najer}, \citenamefont {S\"{o}llner}, \citenamefont {Sekatski}, \citenamefont {Dolique}, \citenamefont {L\"{o}bl}, \citenamefont {Riedel}, \citenamefont {Schott}, \citenamefont {Starosielec}, \citenamefont {Valentin}, \citenamefont {Wieck}, \citenamefont {Sangouard}, \citenamefont {Ludwig},\ and\ \citenamefont {Warburton}}]{Najer2019}%
  \BibitemOpen
  \bibfield  {author} {\bibinfo {author} {\bibfnamefont {D.}~\bibnamefont {Najer}}, \bibinfo {author} {\bibfnamefont {I.}~\bibnamefont {S\"{o}llner}}, \bibinfo {author} {\bibfnamefont {P.}~\bibnamefont {Sekatski}}, \bibinfo {author} {\bibfnamefont {V.}~\bibnamefont {Dolique}}, \bibinfo {author} {\bibfnamefont {M.~C.}\ \bibnamefont {L\"{o}bl}}, \bibinfo {author} {\bibfnamefont {D.}~\bibnamefont {Riedel}}, \bibinfo {author} {\bibfnamefont {R.}~\bibnamefont {Schott}}, \bibinfo {author} {\bibfnamefont {S.}~\bibnamefont {Starosielec}}, \bibinfo {author} {\bibfnamefont {S.~R.}\ \bibnamefont {Valentin}}, \bibinfo {author} {\bibfnamefont {A.~D.}\ \bibnamefont {Wieck}}, \bibinfo {author} {\bibfnamefont {N.}~\bibnamefont {Sangouard}}, \bibinfo {author} {\bibfnamefont {A.}~\bibnamefont {Ludwig}},\ and\ \bibinfo {author} {\bibfnamefont {R.~J.}\ \bibnamefont {Warburton}},\ }\href {https://doi.org/10.1038/s41586-019-1709-y} {\bibfield  {journal} {\bibinfo  {journal} {Nature}\ }\textbf {\bibinfo {volume} {575}},\ \bibinfo
  {pages} {622–627} (\bibinfo {year} {2019})}\BibitemShut {NoStop}%
\bibitem [{\citenamefont {Kaupp}\ \emph {et~al.}(2016)\citenamefont {Kaupp}, \citenamefont {H{\"u}mmer}, \citenamefont {Mader}, \citenamefont {Schlederer}, \citenamefont {Benedikter}, \citenamefont {Haeusser}, \citenamefont {Chang}, \citenamefont {Fedder}, \citenamefont {H{\"a}nsch},\ and\ \citenamefont {Hunger}}]{Kaupp2016}%
  \BibitemOpen
  \bibfield  {author} {\bibinfo {author} {\bibfnamefont {H.}~\bibnamefont {Kaupp}}, \bibinfo {author} {\bibfnamefont {T.}~\bibnamefont {H{\"u}mmer}}, \bibinfo {author} {\bibfnamefont {M.}~\bibnamefont {Mader}}, \bibinfo {author} {\bibfnamefont {B.}~\bibnamefont {Schlederer}}, \bibinfo {author} {\bibfnamefont {J.}~\bibnamefont {Benedikter}}, \bibinfo {author} {\bibfnamefont {P.}~\bibnamefont {Haeusser}}, \bibinfo {author} {\bibfnamefont {H.-C.}\ \bibnamefont {Chang}}, \bibinfo {author} {\bibfnamefont {H.}~\bibnamefont {Fedder}}, \bibinfo {author} {\bibfnamefont {T.~W.}\ \bibnamefont {H{\"a}nsch}},\ and\ \bibinfo {author} {\bibfnamefont {D.}~\bibnamefont {Hunger}},\ }\href {https://doi.org/10.1103/PhysRevApplied.6.054010} {\bibfield  {journal} {\bibinfo  {journal} {Physical Review Applied}\ }\textbf {\bibinfo {volume} {6}},\ \bibinfo {pages} {054010} (\bibinfo {year} {2016})}\BibitemShut {NoStop}%
\bibitem [{\citenamefont {Riedel}\ \emph {et~al.}(2017)\citenamefont {Riedel}, \citenamefont {S{\"o}llner}, \citenamefont {Shields}, \citenamefont {Starosielec}, \citenamefont {Appel}, \citenamefont {Neu}, \citenamefont {Maletinsky},\ and\ \citenamefont {Warburton}}]{Riedel2017}%
  \BibitemOpen
  \bibfield  {author} {\bibinfo {author} {\bibfnamefont {D.}~\bibnamefont {Riedel}}, \bibinfo {author} {\bibfnamefont {I.}~\bibnamefont {S{\"o}llner}}, \bibinfo {author} {\bibfnamefont {B.~J.}\ \bibnamefont {Shields}}, \bibinfo {author} {\bibfnamefont {S.}~\bibnamefont {Starosielec}}, \bibinfo {author} {\bibfnamefont {P.}~\bibnamefont {Appel}}, \bibinfo {author} {\bibfnamefont {E.}~\bibnamefont {Neu}}, \bibinfo {author} {\bibfnamefont {P.}~\bibnamefont {Maletinsky}},\ and\ \bibinfo {author} {\bibfnamefont {R.~J.}\ \bibnamefont {Warburton}},\ }\href {https://doi.org/10.1103/PhysRevX.7.031040} {\bibfield  {journal} {\bibinfo  {journal} {Physical Review X}\ }\textbf {\bibinfo {volume} {7}},\ \bibinfo {pages} {031040} (\bibinfo {year} {2017})}\BibitemShut {NoStop}%
\bibitem [{\citenamefont {Kim}\ \emph {et~al.}(2016{\natexlab{a}})\citenamefont {Kim}, \citenamefont {Richardson}, \citenamefont {Leavitt},\ and\ \citenamefont {Waks}}]{Kim2016TPI}%
  \BibitemOpen
  \bibfield  {author} {\bibinfo {author} {\bibfnamefont {J.-H.}\ \bibnamefont {Kim}}, \bibinfo {author} {\bibfnamefont {C.~J.~K.}\ \bibnamefont {Richardson}}, \bibinfo {author} {\bibfnamefont {R.~P.}\ \bibnamefont {Leavitt}},\ and\ \bibinfo {author} {\bibfnamefont {E.}~\bibnamefont {Waks}},\ }\href {https://doi.org/10.1021/acs.nanolett.6b03295} {\bibfield  {journal} {\bibinfo  {journal} {Nano Letters}\ }\textbf {\bibinfo {volume} {16}},\ \bibinfo {pages} {7061–7066} (\bibinfo {year} {2016}{\natexlab{a}})}\BibitemShut {NoStop}%
\bibitem [{\citenamefont {Grim}\ \emph {et~al.}(2019)\citenamefont {Grim}, \citenamefont {Bracker}, \citenamefont {Zalalutdinov}, \citenamefont {Carter}, \citenamefont {Kozen}, \citenamefont {Kim}, \citenamefont {Kim}, \citenamefont {Mlack}, \citenamefont {Yakes}, \citenamefont {Lee},\ and\ \citenamefont {Gammon}}]{Grim2019}%
  \BibitemOpen
  \bibfield  {author} {\bibinfo {author} {\bibfnamefont {J.~Q.}\ \bibnamefont {Grim}}, \bibinfo {author} {\bibfnamefont {A.~S.}\ \bibnamefont {Bracker}}, \bibinfo {author} {\bibfnamefont {M.}~\bibnamefont {Zalalutdinov}}, \bibinfo {author} {\bibfnamefont {S.~G.}\ \bibnamefont {Carter}}, \bibinfo {author} {\bibfnamefont {A.~C.}\ \bibnamefont {Kozen}}, \bibinfo {author} {\bibfnamefont {M.}~\bibnamefont {Kim}}, \bibinfo {author} {\bibfnamefont {C.~S.}\ \bibnamefont {Kim}}, \bibinfo {author} {\bibfnamefont {J.~T.}\ \bibnamefont {Mlack}}, \bibinfo {author} {\bibfnamefont {M.}~\bibnamefont {Yakes}}, \bibinfo {author} {\bibfnamefont {B.}~\bibnamefont {Lee}},\ and\ \bibinfo {author} {\bibfnamefont {D.}~\bibnamefont {Gammon}},\ }\href {https://doi.org/10.1038/s41563-019-0418-0} {\bibfield  {journal} {\bibinfo  {journal} {Nature Materials}\ }\textbf {\bibinfo {volume} {18}},\ \bibinfo {pages} {963–969} (\bibinfo {year} {2019})}\BibitemShut {NoStop}%
\bibitem [{\citenamefont {Wang}\ \emph {et~al.}(2022{\natexlab{a}})\citenamefont {Wang}, \citenamefont {Zhu}, \citenamefont {Jin}, \citenamefont {Ou}, \citenamefont {Ou},\ and\ \citenamefont {Zhang}}]{wang2022}%
  \BibitemOpen
  \bibfield  {author} {\bibinfo {author} {\bibfnamefont {X.-D.}\ \bibnamefont {Wang}}, \bibinfo {author} {\bibfnamefont {Y.-F.}\ \bibnamefont {Zhu}}, \bibinfo {author} {\bibfnamefont {T.-T.}\ \bibnamefont {Jin}}, \bibinfo {author} {\bibfnamefont {W.-W.}\ \bibnamefont {Ou}}, \bibinfo {author} {\bibfnamefont {X.}~\bibnamefont {Ou}},\ and\ \bibinfo {author} {\bibfnamefont {J.-X.}\ \bibnamefont {Zhang}},\ }\href {https://doi.org/10.1016/j.chip.2022.100018} {\bibfield  {journal} {\bibinfo  {journal} {Chip}\ }\textbf {\bibinfo {volume} {1}},\ \bibinfo {pages} {100018} (\bibinfo {year} {2022}{\natexlab{a}})}\BibitemShut {NoStop}%
\bibitem [{\citenamefont {Uppu}\ \emph {et~al.}(2021)\citenamefont {Uppu}, \citenamefont {Midolo}, \citenamefont {Zhou}, \citenamefont {Carolan},\ and\ \citenamefont {Lodahl}}]{Uppu2021}%
  \BibitemOpen
  \bibfield  {author} {\bibinfo {author} {\bibfnamefont {R.}~\bibnamefont {Uppu}}, \bibinfo {author} {\bibfnamefont {L.}~\bibnamefont {Midolo}}, \bibinfo {author} {\bibfnamefont {X.}~\bibnamefont {Zhou}}, \bibinfo {author} {\bibfnamefont {J.}~\bibnamefont {Carolan}},\ and\ \bibinfo {author} {\bibfnamefont {P.}~\bibnamefont {Lodahl}},\ }\href {https://doi.org/10.1038/s41565-021-00965-6} {\bibfield  {journal} {\bibinfo  {journal} {Nature Nanotechnology}\ }\textbf {\bibinfo {volume} {16}},\ \bibinfo {pages} {1308–1317} (\bibinfo {year} {2021})}\BibitemShut {NoStop}%
\bibitem [{\citenamefont {Papon}\ \emph {et~al.}(2019)\citenamefont {Papon}, \citenamefont {Zhou}, \citenamefont {Thyrrestrup}, \citenamefont {Liu}, \citenamefont {Stobbe}, \citenamefont {Schott}, \citenamefont {Wieck}, \citenamefont {Ludwig}, \citenamefont {Lodahl},\ and\ \citenamefont {Midolo}}]{Papon2019}%
  \BibitemOpen
  \bibfield  {author} {\bibinfo {author} {\bibfnamefont {C.}~\bibnamefont {Papon}}, \bibinfo {author} {\bibfnamefont {X.}~\bibnamefont {Zhou}}, \bibinfo {author} {\bibfnamefont {H.}~\bibnamefont {Thyrrestrup}}, \bibinfo {author} {\bibfnamefont {Z.}~\bibnamefont {Liu}}, \bibinfo {author} {\bibfnamefont {S.}~\bibnamefont {Stobbe}}, \bibinfo {author} {\bibfnamefont {R.}~\bibnamefont {Schott}}, \bibinfo {author} {\bibfnamefont {A.~D.}\ \bibnamefont {Wieck}}, \bibinfo {author} {\bibfnamefont {A.}~\bibnamefont {Ludwig}}, \bibinfo {author} {\bibfnamefont {P.}~\bibnamefont {Lodahl}},\ and\ \bibinfo {author} {\bibfnamefont {L.}~\bibnamefont {Midolo}},\ }\href {https://doi.org/10.1364/optica.6.000524} {\bibfield  {journal} {\bibinfo  {journal} {Optica}\ }\textbf {\bibinfo {volume} {6}},\ \bibinfo {pages} {524} (\bibinfo {year} {2019})}\BibitemShut {NoStop}%
\bibitem [{\citenamefont {Dusanowski}\ \emph {et~al.}(2020)\citenamefont {Dusanowski}, \citenamefont {K{\"o}ck}, \citenamefont {Shin}, \citenamefont {Kwon}, \citenamefont {Schneider},\ and\ \citenamefont {H{\"o}fling}}]{Dusanowski2020}%
  \BibitemOpen
  \bibfield  {author} {\bibinfo {author} {\bibfnamefont {{\L}.}~\bibnamefont {Dusanowski}}, \bibinfo {author} {\bibfnamefont {D.}~\bibnamefont {K{\"o}ck}}, \bibinfo {author} {\bibfnamefont {E.}~\bibnamefont {Shin}}, \bibinfo {author} {\bibfnamefont {S.-H.}\ \bibnamefont {Kwon}}, \bibinfo {author} {\bibfnamefont {C.}~\bibnamefont {Schneider}},\ and\ \bibinfo {author} {\bibfnamefont {S.}~\bibnamefont {H{\"o}fling}},\ }\href {https://doi.org/10.1021/acs.nanolett.0c01771} {\bibfield  {journal} {\bibinfo  {journal} {Nano Letters}\ }\textbf {\bibinfo {volume} {20}},\ \bibinfo {pages} {6357} (\bibinfo {year} {2020})}\BibitemShut {NoStop}%
\bibitem [{\citenamefont {Barclay}\ \emph {et~al.}(2011)\citenamefont {Barclay}, \citenamefont {Fu}, \citenamefont {Santori}, \citenamefont {Faraon},\ and\ \citenamefont {Beausoleil}}]{Barclay2011}%
  \BibitemOpen
  \bibfield  {author} {\bibinfo {author} {\bibfnamefont {P.~E.}\ \bibnamefont {Barclay}}, \bibinfo {author} {\bibfnamefont {K.-M.~C.}\ \bibnamefont {Fu}}, \bibinfo {author} {\bibfnamefont {C.}~\bibnamefont {Santori}}, \bibinfo {author} {\bibfnamefont {A.}~\bibnamefont {Faraon}},\ and\ \bibinfo {author} {\bibfnamefont {R.~G.}\ \bibnamefont {Beausoleil}},\ }\href {https://doi.org/10.1103/PhysRevX.1.011007} {\bibfield  {journal} {\bibinfo  {journal} {Physical Review X}\ }\textbf {\bibinfo {volume} {1}},\ \bibinfo {pages} {011007} (\bibinfo {year} {2011})}\BibitemShut {NoStop}%
\bibitem [{\citenamefont {Kim}\ \emph {et~al.}(2016{\natexlab{b}})\citenamefont {Kim}, \citenamefont {Richardson}, \citenamefont {Leavitt},\ and\ \citenamefont {Waks}}]{Kim2016}%
  \BibitemOpen
  \bibfield  {author} {\bibinfo {author} {\bibfnamefont {J.-H.}\ \bibnamefont {Kim}}, \bibinfo {author} {\bibfnamefont {C.~J.~K.}\ \bibnamefont {Richardson}}, \bibinfo {author} {\bibfnamefont {R.~P.}\ \bibnamefont {Leavitt}},\ and\ \bibinfo {author} {\bibfnamefont {E.}~\bibnamefont {Waks}},\ }\href {https://doi.org/10.1021/acs.nanolett.6b03295} {\bibfield  {journal} {\bibinfo  {journal} {Nano Letters}\ }\textbf {\bibinfo {volume} {16}},\ \bibinfo {pages} {7061–7066} (\bibinfo {year} {2016}{\natexlab{b}})}\BibitemShut {NoStop}%
\bibitem [{\citenamefont {Wang}\ \emph {et~al.}(2022{\natexlab{b}})\citenamefont {Wang}, \citenamefont {Zhu}, \citenamefont {Jin}, \citenamefont {Ou}, \citenamefont {Ou},\ and\ \citenamefont {Zhang}}]{Wangx2022}%
  \BibitemOpen
  \bibfield  {author} {\bibinfo {author} {\bibfnamefont {X.-D.}\ \bibnamefont {Wang}}, \bibinfo {author} {\bibfnamefont {Y.-F.}\ \bibnamefont {Zhu}}, \bibinfo {author} {\bibfnamefont {T.-T.}\ \bibnamefont {Jin}}, \bibinfo {author} {\bibfnamefont {W.-W.}\ \bibnamefont {Ou}}, \bibinfo {author} {\bibfnamefont {X.}~\bibnamefont {Ou}},\ and\ \bibinfo {author} {\bibfnamefont {J.-X.}\ \bibnamefont {Zhang}},\ }\href {https://doi.org/10.1016/j.chip.2022.100018} {\bibfield  {journal} {\bibinfo  {journal} {Chip}\ }\textbf {\bibinfo {volume} {1}},\ \bibinfo {pages} {100018} (\bibinfo {year} {2022}{\natexlab{b}})}\BibitemShut {NoStop}%
\bibitem [{\citenamefont {Jin}\ \emph {et~al.}(2022)\citenamefont {Jin}, \citenamefont {Li}, \citenamefont {Liu}, \citenamefont {Ou}, \citenamefont {Zhu}, \citenamefont {Wang}, \citenamefont {Liu}, \citenamefont {Huo}, \citenamefont {Ou},\ and\ \citenamefont {Zhang}}]{jin2022generation}%
  \BibitemOpen
  \bibfield  {author} {\bibinfo {author} {\bibfnamefont {T.}~\bibnamefont {Jin}}, \bibinfo {author} {\bibfnamefont {X.}~\bibnamefont {Li}}, \bibinfo {author} {\bibfnamefont {R.}~\bibnamefont {Liu}}, \bibinfo {author} {\bibfnamefont {W.}~\bibnamefont {Ou}}, \bibinfo {author} {\bibfnamefont {Y.}~\bibnamefont {Zhu}}, \bibinfo {author} {\bibfnamefont {X.}~\bibnamefont {Wang}}, \bibinfo {author} {\bibfnamefont {J.}~\bibnamefont {Liu}}, \bibinfo {author} {\bibfnamefont {Y.}~\bibnamefont {Huo}}, \bibinfo {author} {\bibfnamefont {X.}~\bibnamefont {Ou}},\ and\ \bibinfo {author} {\bibfnamefont {J.}~\bibnamefont {Zhang}},\ }\href {https://doi.org/10.1021/acs.nanolett.1c03226} {\bibfield  {journal} {\bibinfo  {journal} {Nano Letters}\ }\textbf {\bibinfo {volume} {22}},\ \bibinfo {pages} {586} (\bibinfo {year} {2022})}\BibitemShut {NoStop}%
\bibitem [{\citenamefont {Elshaari}\ \emph {et~al.}(2018)\citenamefont {Elshaari}, \citenamefont {B{\"u}y{\"u}k{\"o}zer}, \citenamefont {Zadeh}, \citenamefont {Lettner}, \citenamefont {Zhao}, \citenamefont {Sch{\"o}ll}, \citenamefont {Gyger}, \citenamefont {Reimer}, \citenamefont {Dalacu}, \citenamefont {Poole}, \citenamefont {J{\"o}ns},\ and\ \citenamefont {Zwiller}}]{elshaari2018strain}%
  \BibitemOpen
  \bibfield  {author} {\bibinfo {author} {\bibfnamefont {A.~W.}\ \bibnamefont {Elshaari}}, \bibinfo {author} {\bibfnamefont {E.}~\bibnamefont {B{\"u}y{\"u}k{\"o}zer}}, \bibinfo {author} {\bibfnamefont {I.~E.}\ \bibnamefont {Zadeh}}, \bibinfo {author} {\bibfnamefont {T.}~\bibnamefont {Lettner}}, \bibinfo {author} {\bibfnamefont {P.}~\bibnamefont {Zhao}}, \bibinfo {author} {\bibfnamefont {E.}~\bibnamefont {Sch{\"o}ll}}, \bibinfo {author} {\bibfnamefont {S.}~\bibnamefont {Gyger}}, \bibinfo {author} {\bibfnamefont {M.~E.}\ \bibnamefont {Reimer}}, \bibinfo {author} {\bibfnamefont {D.}~\bibnamefont {Dalacu}}, \bibinfo {author} {\bibfnamefont {P.~J.}\ \bibnamefont {Poole}}, \bibinfo {author} {\bibfnamefont {K.~D.}\ \bibnamefont {J{\"o}ns}},\ and\ \bibinfo {author} {\bibfnamefont {V.}~\bibnamefont {Zwiller}},\ }\href {https://doi.org/10.1021/acs.nanolett.8b03937} {\bibfield  {journal} {\bibinfo  {journal} {Nano Letters}\ }\textbf {\bibinfo {volume} {18}},\ \bibinfo {pages} {7969} (\bibinfo {year} {2018})}\BibitemShut
  {NoStop}%
\bibitem [{\citenamefont {Osada}\ \emph {et~al.}(2019)\citenamefont {Osada}, \citenamefont {Ota}, \citenamefont {Katsumi}, \citenamefont {Kakuda}, \citenamefont {Iwamoto},\ and\ \citenamefont {Arakawa}}]{Osada2019}%
  \BibitemOpen
  \bibfield  {author} {\bibinfo {author} {\bibfnamefont {A.}~\bibnamefont {Osada}}, \bibinfo {author} {\bibfnamefont {Y.}~\bibnamefont {Ota}}, \bibinfo {author} {\bibfnamefont {R.}~\bibnamefont {Katsumi}}, \bibinfo {author} {\bibfnamefont {M.}~\bibnamefont {Kakuda}}, \bibinfo {author} {\bibfnamefont {S.}~\bibnamefont {Iwamoto}},\ and\ \bibinfo {author} {\bibfnamefont {Y.}~\bibnamefont {Arakawa}},\ }\href {https://doi.org/10.1103/PhysRevApplied.11.024071} {\bibfield  {journal} {\bibinfo  {journal} {Physical Review Applied}\ }\textbf {\bibinfo {volume} {11}},\ \bibinfo {pages} {024071} (\bibinfo {year} {2019})}\BibitemShut {NoStop}%
\bibitem [{\citenamefont {Sun}\ \emph {et~al.}(2010)\citenamefont {Sun}, \citenamefont {Thompson},\ and\ \citenamefont {Nishida}}]{Sun2010}%
  \BibitemOpen
  \bibfield  {author} {\bibinfo {author} {\bibfnamefont {Y.}~\bibnamefont {Sun}}, \bibinfo {author} {\bibfnamefont {S.~E.}\ \bibnamefont {Thompson}},\ and\ \bibinfo {author} {\bibfnamefont {T.}~\bibnamefont {Nishida}},\ }\href {https://doi.org/10.1007/978-1-4419-0552-9} {\emph {\bibinfo {title} {Strain Effect in Semiconductors: Theory and Device Applications}}}\ (\bibinfo  {publisher} {Springer US},\ \bibinfo {year} {2010})\BibitemShut {NoStop}%
\bibitem [{\citenamefont {Lomonte}\ \emph {et~al.}(2021)\citenamefont {Lomonte}, \citenamefont {Wolff}, \citenamefont {Beutel}, \citenamefont {Ferrari}, \citenamefont {Schuck}, \citenamefont {Pernice},\ and\ \citenamefont {Lenzini}}]{lomonte2021}%
  \BibitemOpen
  \bibfield  {author} {\bibinfo {author} {\bibfnamefont {E.}~\bibnamefont {Lomonte}}, \bibinfo {author} {\bibfnamefont {M.~A.}\ \bibnamefont {Wolff}}, \bibinfo {author} {\bibfnamefont {F.}~\bibnamefont {Beutel}}, \bibinfo {author} {\bibfnamefont {S.}~\bibnamefont {Ferrari}}, \bibinfo {author} {\bibfnamefont {C.}~\bibnamefont {Schuck}}, \bibinfo {author} {\bibfnamefont {W.~H.~P.}\ \bibnamefont {Pernice}},\ and\ \bibinfo {author} {\bibfnamefont {F.}~\bibnamefont {Lenzini}},\ }\href {https://doi.org/10.1038/s41467-021-27205-8} {\bibfield  {journal} {\bibinfo  {journal} {Nature Communications}\ }\textbf {\bibinfo {volume} {12}},\ \bibinfo {pages} {6847} (\bibinfo {year} {2021})}\BibitemShut {NoStop}%
\bibitem [{\citenamefont {Zhai}\ \emph {et~al.}(2020)\citenamefont {Zhai}, \citenamefont {L{\"o}bl}, \citenamefont {Nguyen}, \citenamefont {Ritzmann}, \citenamefont {Javadi}, \citenamefont {Spinnler}, \citenamefont {Wieck}, \citenamefont {Ludwig},\ and\ \citenamefont {Warburton}}]{Zhai2020}%
  \BibitemOpen
  \bibfield  {author} {\bibinfo {author} {\bibfnamefont {L.}~\bibnamefont {Zhai}}, \bibinfo {author} {\bibfnamefont {M.~C.}\ \bibnamefont {L{\"o}bl}}, \bibinfo {author} {\bibfnamefont {G.~N.}\ \bibnamefont {Nguyen}}, \bibinfo {author} {\bibfnamefont {J.}~\bibnamefont {Ritzmann}}, \bibinfo {author} {\bibfnamefont {A.}~\bibnamefont {Javadi}}, \bibinfo {author} {\bibfnamefont {C.}~\bibnamefont {Spinnler}}, \bibinfo {author} {\bibfnamefont {A.~D.}\ \bibnamefont {Wieck}}, \bibinfo {author} {\bibfnamefont {A.}~\bibnamefont {Ludwig}},\ and\ \bibinfo {author} {\bibfnamefont {R.~J.}\ \bibnamefont {Warburton}},\ }\href {https://doi.org/10.1038/s41467-020-18625-z} {\bibfield  {journal} {\bibinfo  {journal} {Nature Communications}\ }\textbf {\bibinfo {volume} {11}},\ \bibinfo {pages} {4745} (\bibinfo {year} {2020})}\BibitemShut {NoStop}%
\end{thebibliography}%
\end{document}